# Twin Boundary Mediated Flexoelectricity in LaAlO$_3$

Christopher A. Mizzi, Binghao Guo, Laurence D. Marks

Department of Materials Science and Engineering, Northwestern University, Evanston IL, 60208, USA

**Abstract**

Flexoelectricity has garnered much attention owing to its ability to bring electromechanical functionality to non-piezoelectric materials and its nanoscale significance. In order to move towards a more complete understanding of this phenomenon and improve the efficacy of flexoelectric-based devices, it is necessary to quantify microstructural contributions to flexoelectricity. Here we characterize the flexoelectric response of bulk centrosymmetric LaAlO$_3$ crystals with different twin boundary microstructures. We show that twin boundary flexoelectric contributions are comparable to intrinsic contributions at room temperature and enhance the flexoelectric response by ~4x at elevated temperatures. Additionally, we observe time-dependent and non-linear flexoelectric responses associated with strain-gradient-induced twin boundary polarization. These results are explained by considering the interplay between twin boundary orientation, beam-bending strain fields, and pinning site interactions.

Inducing an electrical response from a mechanical stimulus (or a mechanical response from an electrical stimulus) in insulators is highly desirable for actuation, sensing, and energy harvesting applications. Historically, such electromechanical functionality has been derived from piezoelectricity (the coupling of strain and polarization), but piezoelectricity faces a fundamental limitation: as a bulk property it only exists in non-centrosymmetric materials [1]. This, coupled with the prevalence of lead in common piezoelectrics, poses a significant materials selection challenge [2].

One approach to overcome these issues is to replace piezoelectric materials with flexoelectric (FxE) materials. Flexoelectricity (the coupling of strain gradient and polarization) allows for the mechanical polarization of all insulators, even those with centrosymmetric space groups, because applying a strain gradient inherently breaks inversion symmetry [3,4]. Since Ma and Cross discovered large FxE responses in relaxor ferroelectric ceramics [5], research on flexoelectricity in oxides has flourished [3,4]. However, an abundance of fundamental questions persists. Among the most pressing relates to the role of extrinsic contributions to flexoelectricity (e.g. terms related to microstructure, point defects, etc.). Extrinsic contributions are bound to be prevalent in ceramics utilized for electromechanical applications and recent work has shown they can dominate intrinsic (i.e. pure crystallographic) FxE contributions: the overall FxE response has sizeable modifications from free carriers in semiconductors [6], polar nano-regions in relaxor ferroelectrics [7,8], and polar selvedge regions in ferroelectric ceramics [9]. Identifying and separating extrinsic and intrinsic contributions to flexoelectricity is an important step towards improving our fundamental understanding of flexoelectricity, closing the sizable divide between the experimental and theoretical [10-14] states of the field, and substantiating the viability of flexoelectricity for practical applications.

Oxides which, when grown at high temperatures are single crystals, and then develop twin boundaries (TBs) on cooling are well-poised to elucidate microstructural FxE contributions because they bridge the gap between single crystals and polycrystalline ceramics. While experiments on such oxides, like $SrTiO_3$ (STO) [15], have indicated TB contributions to flexoelectricity are important, there has been no quantitative analysis of TB FxE contributions, i.e. determining the FxE coefficient of a TB. Moreover, the role of TB microstructure in the overall FxE response of a sample has not been addressed. LAO is a rhombohedral "332" perovskite with space group $R\bar{3}c$ at room temperature and atmospheric pressure [16] which twins due to an improper ferroelastic phase transition at 550 °C [17]. It is an ideal material to examine TB effects on flexoelectricity because it is twinned at room temperature and the crystallography of its TBs [18-22], as well as their elastic response to dynamic mechanical stimuli [23-26], has been extensively studied. Understanding flexoelectricity in LAO is also important because it is commonly used as a substrate and film in thin-film growth [27,28] (large FxE responses occur at length scales relevant to thin films because of the intrinsic size scaling of strain gradients [29-31]) and there has been recent interest in utilizing FxE couplings to modify the two-dimensional electron gas at the LAO/STO interface [32-35]. More generally, degrees of freedom often vary rapidly over short distances in the vicinity of domain walls, such as TBs. Consequently, couplings involving gradients, such as flexoelectricity, can be particularly potent in ferroic materials [36-39], leading to novel properties that are otherwise forbidden in the bulk of the material [40], such as static TB polarization. While the evidence for static TB polarization in LAO [41-43] is compelling, it is unclear how, quantitatively, these static polarizations impact macroscopic measurements or evolve under applied strain gradients.

In this Letter we show that extrinsic FxE contributions associated with TBs are of the same order as intrinsic contributions to flexoelectricity in LAO. Using a low-frequency beam-bending method we measure the effective FxE coefficient in {100}$_{pseudo-cubic}$ LAO crystals with different TB microstructures, demonstrating that strain-gradient-induced TB polarization can yield anelastic and non-linear FxE responses, and at elevated temperatures, enhance the effective FxE coefficient of LAO by ~4x its single crystal value. These enhancements are explained by considering the interplay between TB orientation, beam-bending strain fields, and pinning site interactions.

Throughout this work FxE characterization was performed following Zubko et al. [15]: a sample cut into a beam geometry was subjected to low-frequency bending by a dynamic mechanical analyzer while the short-circuit current generated by the FxE effect was measured with a lock-in amplifier (Fig. 1(a), and Supplemental Materials for details [44]). This measurement method yields an effective FxE coefficient $\mu_{eff}$ that is a linear combination of FxE coefficient tensor components [15]. As derived in the Supplemental Materials [44], the polarization measured from a cubic material with {100} faces is

$$P_z = (\mu_{zxxz}(1-\nu) - \mu_{zzzz}\nu) = \mu_{eff}\epsilon_{xx,z} \qquad (1)$$

where $P_z$ is the component of the polarization along the measurement direction, $\mu_{ijkl}$ are FxE coefficient tensor components, $\epsilon_{ij,k}$ is the gradient of strain $\epsilon_{ij}$ with respect to the $k$ coordinate, and $\nu$ is the relevant Poisson's ratio. The deviation from the cubic perovskite structure is small in LAO; this approximation is addressed in the Supplemental Materials [44].

To examine microstructural contributions to flexoelectricity in LAO, it is necessary to isolate TB effects from crystallographic ones. To this end, we solely focus on {100}$_{pseudo-cubic}$ crystals and begin with the FxE characterization of a LAO crystal without twins, which was cut from a twin-free portion of a larger crystal. As shown in Fig. 1(b), TB-free LAO exhibits a linear

FxE response with $\mu_{eff}$ = 3.2±0.3 nC/m. Using the dielectric constant of LAO [17], this corresponds to a flexocoupling voltage of 14.5±1.4 V. This value is in agreement with a recent measurement of the FxE response of LAO using a different approach [35].

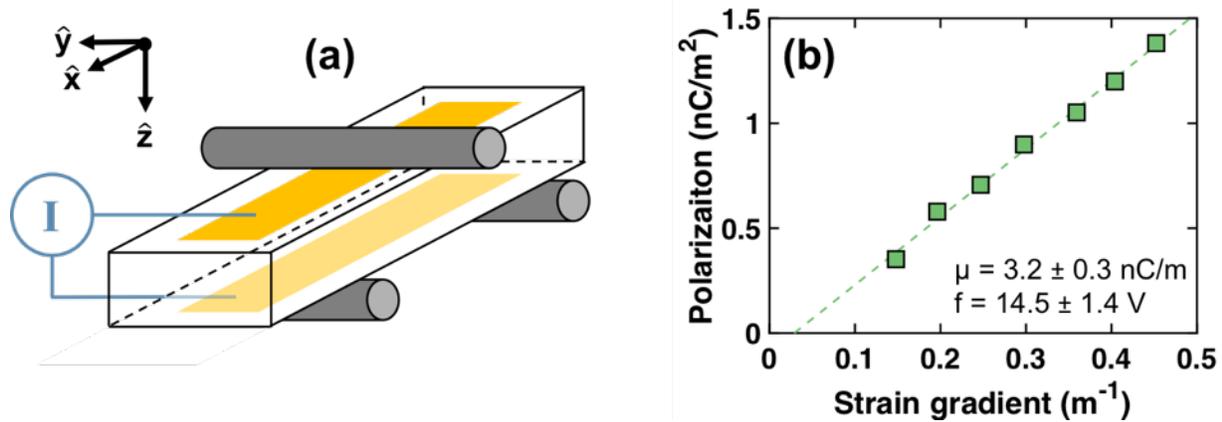

Fig 1. (a) Overview of experimental flexoelectric characterization. The short-circuit current induced by low-frequency three-point bending is measured from a sample cut into a beam geometry. (b) Flexoelectric characterization of a LaAlO$_3$ crystal with no twin boundaries. The strain-gradient-induced polarization in this sample is linear, with an effective flexoelectric coefficient of 3.2±0.3 nC/m (flexocoupling voltage of 14.5±1.4 V). The linear fit is shown as a dashed line. The uncertainty corresponds to the 95% confidence interval of the fit.

Having established the FxE response of LAO with no twins, TB contributions were studied by measuring flexoelectricity in two samples with different lamellar microstructures. The TB orientations in each sample are described in Fig. 2 and they will henceforth be referred to as Type I and Type II. As shown in Fig. 2, the initial FxE response of the Type I sample was linear with $\mu_{eff}$ = 3.6±0.2 nC/m. Additional measurements were then performed while increasing and decreasing the strain gradient (strain gradient cycling). Throughout these measurements the FxE response remained linear and $\mu_{eff}$ increased, eventually reaching a steady-state value of 4.8±0.3 nC/m. Measurements performed the next day on the same sample indicated a partial recovery of the initial FxE response, followed by a return to the same steady-state $\mu_{eff}$ after strain gradient cycling (Supplemental Materials [44]). The Type II sample exhibited qualitatively similar behavior: $\mu_{eff}$ increased from 3.1±0.3 nC/m to a steady-state value of 3.5±0.2 nC/m after strain gradient cycling, remaining linear throughout the measurements. In both Type I and II samples, post-experiment imaging indicated no permanent changes to TB microstructure (Supplemental Materials [44]).

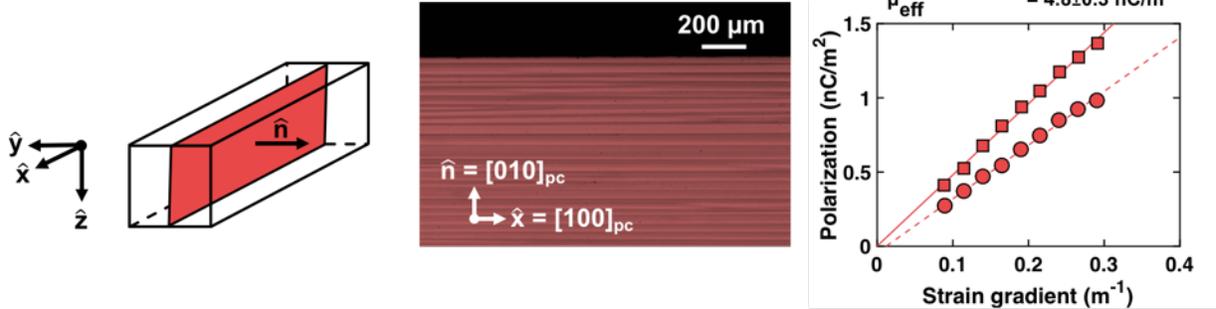

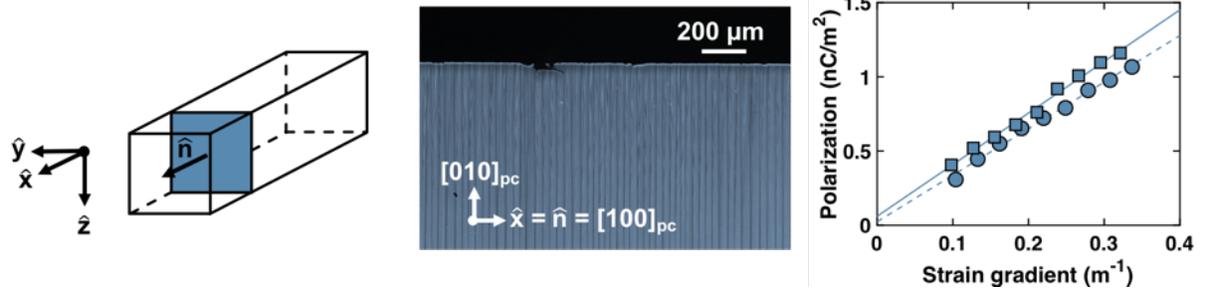

Fig 2. Flexoelectric characterization of LaAlO$_3$ crystals with uniform, lamellar twin boundary microstructures. (a) Type I boundaries have a normal ($\hat{n}$) perpendicular to the long axis of the sample ($\hat{x}$) as shown by reflection polarized optical microscopy. The flexoelectric response of the Type I sample was initially linear and increased to an effective steady-state flexoelectric coefficient of 4.8±0.3 nC/m (flexocoupling voltage of 21.7±1.4 V) after strain gradient cycling. (b) Type II boundaries have a normal parallel to the long axis of the sample as shown by reflection polarized optical microscopy. The flexoelectric response of the Type II sample was initially linear and increased to an effective steady-state flexoelectric coefficient of 3.5±0.2 nC/m (flexocoupling voltage of 15.8±0.9 V) after strain gradient cycling. In both plots, lines are linear fits with circles/dashed lines indicating initial measurements and squares/solid lines indicating steady-state measurements after strain gradient cycling. Uncertainties correspond to the 95% confidence interval of the fit.

These results suggest TB contributions to flexoelectricity are anelastic (time-dependent and elastic) and dependent on TB orientation. The former is a natural consequence of anelastic TB deformation [23-26,45], and the latter is to be expected if TBs in LAO couple to the measured FxE response. Assuming TBs can be polarized by strain gradients, their orientation dictates which strain gradient components couple to the measured polarization. For example, Type I TBs are parallel to the x-z plane so FxE contributions to $P_z$ can only arise from $\epsilon_{xx,z}$ and $\epsilon_{zz,z}$. As shown in the Supplemental Materials [44], the presence of Type I and II TBs modifies $\mu_{eff}$ according to

$$\mu_{eff}^{I} = \mu_{eff}^{bulk} + \mu_{zxxz}^{TB} - \nu \mu_{zzzz}^{TB} \qquad (2)$$

$$\mu_{eff}^{II} = \mu_{eff}^{bulk} - \nu (\mu_{zxxz}^{TB} + \mu_{zzzz}^{TB}) \qquad (3)$$

where $\mu_{eff}^{I,II}$ are $\mu_{eff}$ for the Type I and II samples, $\mu_{eff}^{bulk}$ is $\mu_{eff}$ for the TB-free sample, $\nu$ is the Poisson's ratio, and $\mu_{ijkl}^{TB}$ are TB FxE coefficient tensor components. From Eq. (2) and (3), measurements on the TB-free, Type I, and Type II samples, and the literature Poisson's ratio [46], $\mu_{zxxz}^{TB} = 0.9$ nC/m and $\mu_{zzzz}^{TB} = -1.9$ nC/m. Three comments on this result: (1) the two-dimensional nature of TBs allows for an extraction of individual TB tensor components unlike the case of bulk tensor components which cannot be determined from beam-bending experiments alone [15]; (2) while there is a growing body of experimental [42,43,47,48] and theoretical [41,49,50] work in LAO and similar bulk centrosymmetric perovskites indicating the presence of polar TBs, the experiments reported in this Letter only measure changes in polarization and, as conducted, are insensitive to intrinsic polarization; (3) the local TB polarization associated with these measurements can be estimated as $P_{local}^{TB} \approx \frac{\mu^{TB}}{w} \sim 0.5$ C/m² where the TB width $w$ is taken to be 2 nm [51]. This value is comparable to local polarization measurements of TB in CaTiO$_3$ [47] and dislocation cores in STO [52].

Together these results provide evidence that TBs play a dominant role in determining the FxE response of twinned materials. To validate that these effects can be genuinely attributed to TBs, three additional experiments were performed on samples with a mixture of TB orientations (Supplemental Materials for images [44]). In the first, $\mu_{eff}$ was measured as a function of temperature. It is established that increasing temperature provides thermal energy for TBs to escape pinning sites [23-26,45]: at first increasing temperature locally enhances elastic TB deformations, but after a sample-dependent threshold (typically ~100 °C), large-scale TB motion and annihilation become possible. The temperature dependence shown in Fig. 3(a) correlates well with these trends. Namely, $\mu_{eff}$ increases between ~30-100 °C as TBs become more deformable and decreases once TB motion becomes possible, with a maximum corresponding to the onset of motion. We note that unlike the room temperature observations, increasing temperature permanently changed the microstructure (Supplemental Materials [44]) and the FxE response of twinned LAO at elevated temperatures was found to be very sensitive to the initial TB microstructure.

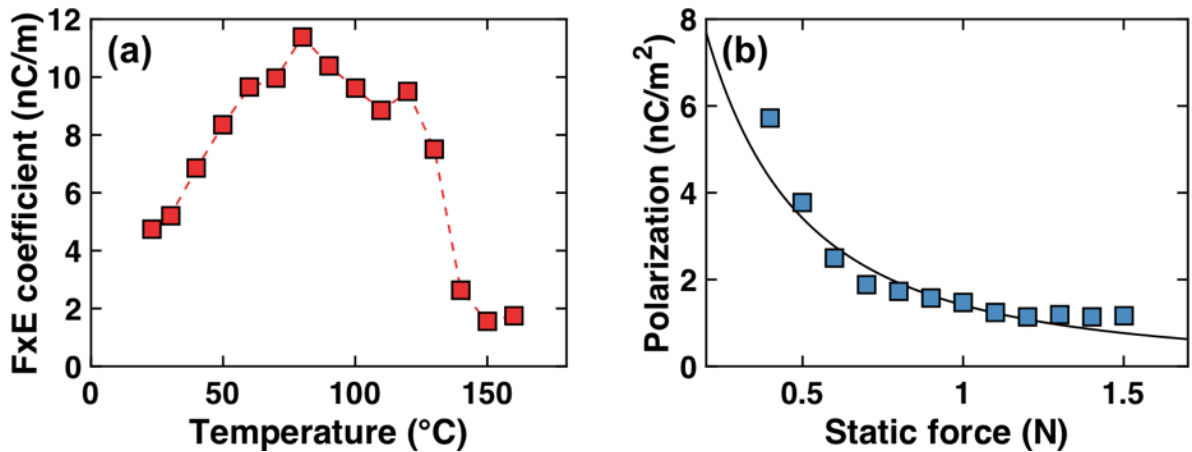

Fig 3. Flexoelectric measurements in LaAlO$_3$ crystals with a mixture of twin boundary orientations. (a) The temperature dependence of the flexoelectric coefficient correlates with the temperature dependence of twin boundary motion. Dashed lines are visual guides. (b) Flexoelectric polarization at room temperature and fixed dynamic force decreases with increasing static force because of the non-linear elastic response of twin boundaries. Solid line is a fit using a classic catenary cable model.

The second experiment to confirm our interpretation was measuring the FxE polarization at a constant dynamic force while varying the static force holding the sample in place during three-point bending. In typical single crystals the FxE polarization is proportional to dynamic force and insensitive to static force, whereas in twinned crystals, such as low temperature STO [15], FxE polarization can depend on static force. We attribute the decrease in FxE polarization with increasing static force to the non-linear relationship between the deformation of a pinned TB and the applied force. As shown in the Supplemental Materials [44], this static force dependency is well-captured by modeling the deformation-force relationship of a pinned TB as a classic catenary cable.

The last experiment was measuring the FxE response of a sample with a mixture of Type I and II TBs at fixed static force (comparable to the results in Fig. 1 and 2). As shown in Fig. 4, $\mu_{eff}$ anelastically increased from an initial value of 3.6±0.1 nC/m to a steady-state value of 4.0±0.2 nC/m (as determined from the first 5 data points in each data set) after strain gradient cycling. This anelasticity is similar to that of the pure Type I and II samples and the steady-state $\mu_{eff}$ is between that of the pure Type I and Type II samples, which is consistent with the mixture of Type I and II TBs present in the sample. Additionally, these measurements indicate a non-linear FxE response when the strain gradient exceeded ~0.25 m$^{-1}$. This starkly contrasts with the FxE responses of the other samples which remained linear for all strain gradients used in these experiments. The non-linearity became more pronounced in steady-state, but a linear FxE response was always recovered by decreasing the strain gradient below ~0.25 m$^{-1}$. Given the return to linear flexoelectricity below a certain strain gradient and post-experiment imaging confirming no permanent changes to TB microstructure (Supplemental Materials [44]), the non-linearity is attributed to TB-pinning site interactions. This behavior is also consistent with the catenary cable model described in the Supplemental Materials [44], further supporting our interpretation.

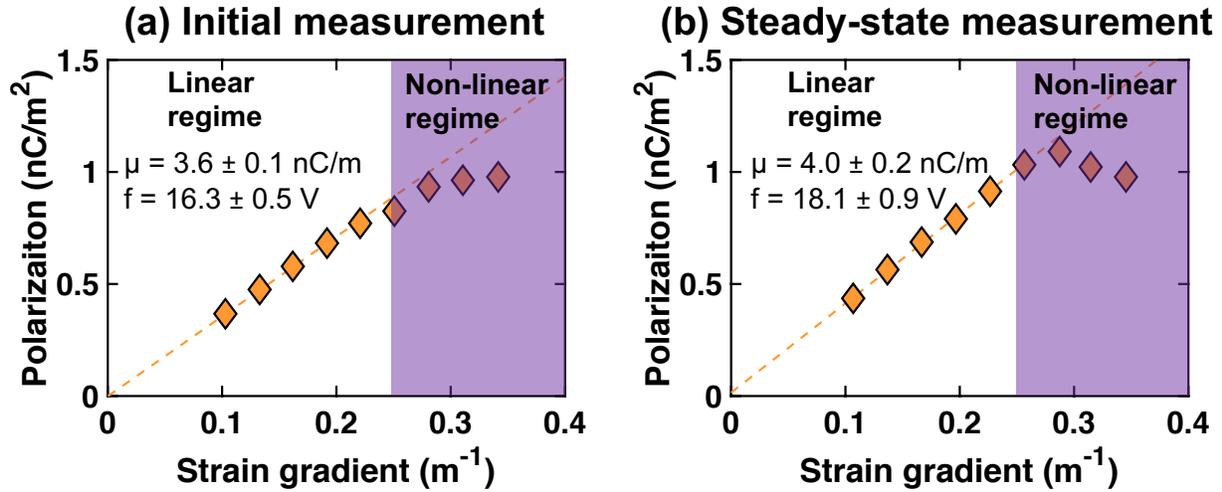

Fig 4. Flexoelectric characterization of a LaAlO$_3$ crystal with a mixture of Type I and II twin boundary orientations. (a) The initial flexoelectric response is linear at low strain gradients with an effective flexoelectric coefficient of 3.6±0.1 nC/m (flexocoupling voltage of 16.3±0.5 V), and non-linear above a strain gradient of ~0.25 m$^{-1}$. (b) The steady-state flexoelectric response after strain gradient cycling remains linear at low strain gradients with an increased effective flexoelectric coefficient of 4.0±0.2 nC/m (flexocoupling voltage of 18.1±0.9 V). There is pronounced non-linear behavior above ~0.25 m$^{-1}$.

In summary we have found that extrinsic contributions to flexoelectricity originating from strain-gradient-induced TB polarization can be substantial and surpass intrinsic contributions to flexoelectricity in twinned LAO. This experimentally confirms the importance of microstructure in FxE phenomena. The FxE characterization of twin-free LAO crystals suggests that the intrinsic flexocoupling voltage is ~14 V, which is enhanced by TB movement at elevated temperature to ~55 V. TB FxE contributions are found to be very sensitive to the details of TB microstructure and responsible for previously unobserved anelastic and non-linear flexoelectricity at room temperature.

**Acknowledgements**

This work was supported by the U.S. Department of Energy, Office of Science, Basic Energy Sciences, under Award No. DE-FG02-01ER45945. CAM and BG performed sample preparation

FxE characterization, and optical microscopy supervised by LDM. CAM performed the analysis supervised by LDM. All authors contributed to the writing of the paper.

**Supplemental Material: Twin Boundary Mediated Flexoelectricity in LaAlO$_3$**

Christopher A. Mizzi, Binghao Guo, Laurence D. Marks[1]

Department of Materials Science and Engineering, Northwestern University, Evanston, IL 60208, USA

**Corresponding Author**

[1]To whom correspondence should be addressed. Email: L-marks@northwestern.edu.

**S1. Flexoelectric characterization overview**

Commercially available 0.5 mm thick crystalline substrates (MTI Corporation) were cut into 10 mm × 3 mm samples. After the cut samples were sonicated in acetone and then ethanol, a sputter coater was used to deposit ~50 nm thick, 8 mm × 2 mm gold electrodes on each 10 mm x 3 mm surface. Copper wires were attached to the electrodes with silver paste and the samples were baked at 300 °C for 2 hours to improve mechanical stability and electrical conductivity.

FxE characterization was performed using a three-point bending method similar to the one described by Zubko et al. [1]. A TA Instruments RSA-III dynamic mechanical analyzer (DMA) was used to bend samples at 33 Hz while a Signal Recovery 7265 Dual Phase DSP lock-in amplifier measured the short-circuit current generated due to the FxE effect. Unless otherwise stated, a constant static force of ~1.5 N was used to hold the sample in place while an oscillatory force was applied. The displacement induced in the sample by the oscillatory force was used to calculate the average strain gradient across the electrode area which, according to the Euler-Bernoulli beam equation [1-3], is given by

$$\bar{\epsilon}_{xx,z} = \overline{\frac{\partial \epsilon_{xx}}{\partial z}} = 12 \frac{L-a}{L^3} u_z \qquad (1)$$

where $L$ is the distance between the three-point bending supports, $a$ is the length of the electrode, and $u_z$ is the displacement at the center of the beam in the $\hat{z}$ direction. Using the sample dimensions and the typical strain gradients (~0.1 m$^{-1}$) calculated from Eq. (1), the Searle parameter ($\frac{w^2}{tR}$, where $w$ is the sample width, $t$ is the sample thickness, and $R$ is the radius of curvature) is ~1e-3 which firmly places these samples in the beam bending regime as opposed to the plate bending regime [4,5].

The average polarization generated along the bending direction via the FxE effect is calculated from the measured current via

$$|\bar{P}_z| = \frac{I}{2\omega A} \qquad (2)$$

where $|\bar{P}_z|$ is the magnitude of the average polarization along the bending direction, $I$ is the current measured by the lock-in amplifier, $\omega$ is the radial frequency of the oscillatory force, and $A$ is the electrode area.

Strain gradients and polarizations shown in FxE plots in this paper correspond to values calculated according to Eq. (1) and (2). Each data point is an average of ~800 measurements. Throughout these experiments the FxE response of each sample was measured under increasing and decreasing strain gradients (strain gradient cycling).

For temperature-dependent FxE characterization, the temperature was controlled using a forced-air convection oven on the DMA. The temperature was changed in small increments (<5 °C) and the sample was allowed to equilibrate for ~5 minutes before the FxE measurements. Each data point shown in the temperature-dependent data corresponds to the slope of a line made from measurements of at least 5 strain gradient values, each corresponding to the average of ~800 measurements.

## S2. Validity of approximating LaAlO$_3$ as a cubic perovskite

Although LAO is a rhombohedral perovskite at room temperature and atmospheric pressure, the rhombohedral distortions are small [6], so LAO can be approximated as a cubic material. To assess the validity of this approximation, we utilize arguments based upon the first-principles theory of flexoelectricity [7,8]. Namely, we compare how terms that impact the electronic FxE contributions (i.e. dynamical octupole moments and high frequency dielectric constant) and lattice-dipole FxE contributions (i.e. Born charges, static dielectric constant, and force-constant matrix) differ in the rhombohedral and cubic LAO phases.

First, we consider the electronic FxE contributions: the high-frequency dielectric constants in the two phases are nearly identical [9] and our unpublished test calculations indicate the dynamical octupole moment is largely unaffected by changes in local crystal structure. Even if there are some changes in the dynamical octupole moment, electronic contributions tend to be overshadowed by lattice contributions in the effective FxE coefficient owing to the large static dielectric constant.

Next, we consider the lattice-dipole FxE contributions. The Born charges differ by <3% in both phases [9], indicating that the dynamical dipole moments should be very similar in the rhombohedral and cubic phases. Also, the static dielectric constant changes by ~10% [9], so converting from fixed D to fixed E boundary conditions will involve scaling by similar terms. The most likely deviation from the cubic FxE response relates to the force-constant matrix: there are more off-diagonal terms in the rhombohedral phase so the lattice-dipole mode will involve more terms. Since the second-order force moments are essentially lattice-resolved elastic constants [8], it is possible to use differences in elastic constants as a proxy for differences in the lattice contributions to the FxE response in the two phases. Since the lattice is more anisotropic in the rhombohedral phase compared to the cubic phase [10] this has potentially the largest impact on the overall FxE response. However, for the purposes of understanding TB contributions to flexoelectricity we argue the cubic approximation is justified.

**S3. Effective flexoelectric coefficient in cubic single crystal with {100} faces**

Using this measurement method, an effective FxE coefficient defined as

$$P_z = \mu_{eff} \epsilon_{xx,z} \tag{3}$$

is measured. For notational simplicity, the average symbols are not used from this point on. This effective FxE coefficient is a linear combination of tensor components. Below we derive the expression for $\mu_{eff}$ for a single crystal cubic material cut into a beam geometry with <100> axes.

First, it is necessary to define the strain fields induced by beam-bending. As derived in Landau and Lifshitz [11], the beam-bending strain tensor for a cubic material under the assumption of no body forces, which is valid for small polarizations where the elastic strain energy dominates the polarization energy, is given by:

$$\epsilon_{ij} = \frac{z}{R}\begin{pmatrix} 1 & 0 & 0 \\ 0 & -\nu & 0 \\ 0 & 0 & -\nu \end{pmatrix} \tag{4}$$

In this expression, $z$ is the distance from the neutral plane, $R$ is the radius of curvature, and $\nu = \frac{C_{1122}}{C_{1111}+C_{1122}}$ is the relevant Poisson coefficient. Note, the axes are defined according to Figures 1 and 2 of the main text.

This strain tensor yields three non-zero strain gradients

$$\epsilon_{xx,z} = \frac{\partial \epsilon_{xx}}{\partial z} = \frac{1}{R} \tag{5}$$

$$\epsilon_{yy,z} = \frac{\partial \epsilon_{yy}}{\partial z} = -\frac{\nu}{R} = -\nu\epsilon_{xx,z} \tag{6}$$

$$\epsilon_{zz,z} = \frac{\partial \epsilon_{zz}}{\partial z} = -\frac{\nu}{R} = \epsilon_{yy,z} = -\nu\epsilon_{xx,z} \tag{7}$$

Substituting the non-zero strain gradient tensor components into the constitutive FxE equation for a cubic material where the polarization is measured in the z direction yields

$$P_z = \mu_{zxxz}\ \epsilon_{xx,z} + \mu_{zyyz}\ \epsilon_{yy,z} + \mu_{zzzz}\ \epsilon_{zz,z} \tag{8}$$

This can be simplified by using Eq. (6) and (7).

$$P_z = \left(\mu_{zxxz} - \mu_{zyyz}\ \nu - \mu_{zzzz}\ \nu\right)\epsilon_{xx,z} \tag{9}$$

Furthermore, since we have assumed a cubic crystal with x=y=z=<100>

$$\mu_{zxxz} = \mu_{zyyz} \tag{10}$$

which leads to the simplification that

$$P_z = \left(\mu_{zxxz}\ (1-\nu) - \mu_{zzzz}\ \nu\right)\epsilon_{xx,z} \tag{11}$$

This provides an expression for the measured $\mu_{eff}$

$$\mu_{eff} = -\nu\ \mu_{zzzz} + (1-\nu)\ \mu_{zxxz} \tag{12}$$

that applies to the measurements on the TB-free LAO shown in Figure 1 of the main text.

**S4. Effective flexoelectric coefficient in Type I and II twinned cubic crystals**

When TBs are present in a crystal it is possible for them to develop a polarization in response to the strain gradients applied during beam bending. This means the total polarization measured in low-frequency beam bending experiments has two contributions

$$P_z = P_z^{bulk} + P_z^{TB} \tag{13}$$

where $P_z$ is the total FxE polarization in a sample with TBs, $P_z^{bulk}$ is the contribution arising from the sample bulk (and surface [12,13]), and $P_z^{TB}$ is the contribution arising from TBs. From this starting point, we derive expressions for $\mu_{eff}^{I}$ and $\mu_{eff}^{II}$ which correspond to effective FxE coefficients measured for samples solely containing Type I and II TBs. Type I and II TBs are defined in Figure 2 of the main text. $\mu_{eff}^{I}$ and $\mu_{eff}^{II}$ are defined according to Eq. (14) and (15) below.

$$P_z^I = \mu_{eff}^I \epsilon_{xx,z} \tag{14}$$

$$P_z^{II} = \mu_{eff}^{II} \epsilon_{xx,z} \tag{15}$$

Since all crystals considered here have the same crystallographic orientation, $P_z^{bulk}$ will be given by Eq. (11) with $\mu_{eff}^{bulk}$ given by Eq. (12). Therefore, we solely focus on $P_z^{TB}$ from Eq. (13).

Type I TBs are parallel to the x-z plane, so contributions to the polarization from Type I TBs, $P_z^{TB,I}$, can only couple to $\epsilon_{xx,z}$ and $\epsilon_{zz,z}$. This is expressed by

$$P_z^{TB,I} = \mu_{zxxz}^{TB} \epsilon_{xx,z} + \mu_{zzzz}^{TB} \epsilon_{zz,z} \tag{16}$$

Similarly, Type II TBs are parallel to the y-z plane, so contributions to the polarization from Type II TBs, $P_z^{TB,II}$, can only couple to $\epsilon_{yy,z}$ and $\epsilon_{zz,z}$. Therefore,

$$P_z^{TB,II} = \mu_{zyyz}^{TB} \epsilon_{yy,z} + \mu_{zzzz}^{TB} \epsilon_{zz,z} \tag{17}$$

Both Eq. (16) and (17) can be simplified using Eq. (6), (7), and (10).

$$P_z^{TB,I} = (\mu_{zxxz}^{TB} - \nu \mu_{zzzz}^{TB})\epsilon_{xx,z} \tag{18}$$

$$P_z^{TB,II} = -\nu (\mu_{zxxz}^{TB} + \mu_{zzzz}^{TB})\epsilon_{xx,z} \tag{19}$$

After reintroducing the bulk contributions, we have:

$$P_z^I = P_z^{bulk} + P_z^{TB,I} = \left(\mu_{eff}^{bulk} + \mu_{zxxz}^{TB} - \nu \mu_{zzzz}^{TB}\right)\epsilon_{xx,z} \tag{20}$$

$$P_z^{II} = P_z^{bulk} + P_z^{TB,II} = \left(\mu_{eff}^{bulk} - \nu (\mu_{zxxz}^{TB} + \mu_{zzzz}^{TB})\right)\epsilon_{xx,z} \tag{21}$$

yielding the following expressions for $\mu_{eff}^I$ and $\mu_{eff}^{II}$.

$$\mu_{eff}^I = \mu_{eff}^{bulk} + \mu_{zxxz}^{TB} - \nu \mu_{zzzz}^{TB} \tag{22}$$

$$\mu_{eff}^{II} = \mu_{eff}^{bulk} - \nu (\mu_{zxxz}^{TB} + \mu_{zzzz}^{TB}) \tag{23}$$

Some comments on assumptions:

- In writing Eq. (18) and (19) we have assumed that the TBs with different orientations will be described by the same FxE coefficient tensor.

- We assume the bulk Poisson's ratio applies to the TB.

- We assume the same x=y symmetry is upheld in the TB so that $\mu^{TB}_{zxxz} = \mu^{TB}_{zyyz}$.

## S5. Origin of non-linear flexoelectricity and static force dependence

Fundamentally, FxE measurements measure changes in polarization arising from changes in strain gradient. Working in one-dimension, this is expressed as

$$\frac{dP}{dt} = \mu \frac{d\kappa}{dt} \qquad (24)$$

where $\frac{dP}{dt}$ is the change in polarization arising from a change in strain gradient $\frac{d\kappa}{dt}$ owing to a FxE coupling whose strength is determined by the FxE coefficient $\mu$. Since changes in strain gradient are experimentally controlled by changes in the applied force, it is advantageous to expand Eq. (24) in terms of the static force $f_S$ and dynamic force $f_D$.

$$\frac{dP}{dt} = \mu \left( \frac{\partial \kappa}{\partial f_S} \frac{df_S}{dt} + \frac{\partial \kappa}{\partial f_D} \frac{df_D}{dt} \right) = \mu \frac{\partial \kappa}{\partial f_D} \frac{df_D}{dt} \qquad (25)$$

In Eq. (25) we have made use of the fact that $f_S$ is a constant. It is also convenient to expand Eq. (25) in terms of displacement

$$\frac{dP}{dt} = \mu \frac{\partial \kappa}{\partial u} \frac{\partial u}{\partial f_D} \frac{df_D}{dt} \qquad (26)$$

because the strain gradient of a bent feature (e.g. beam, line defect, planar defect) is proportional the displacement of its center $u$ from its unbent reference state. Writing

$$\kappa = C\, u \qquad (27)$$

where C is a geometric/elastic constant and substituting into Eq. (26) yields

$$\frac{dP}{dt} = C\, \mu\, \frac{\partial u}{\partial f_D} \frac{df_D}{dt} \qquad (28)$$

From this equation, it is apparent that a non-linear elastic material (i.e. a material exhibiting a non-linear response between $u$ and $f$) will yield a non-linear FxE response.

In the case of a pinned defect, the deformation response is similar to the classic catenary cable problem, with the gravitational force replaced by the applied force, see Figure S1. As such, the displacement $u(x)$ of a pinned defect between two pinning sites is given by

$$u(x) = \frac{T}{f_S + f_D} \cosh\left(\frac{f_S + f_D}{T} x\right) \tag{29}$$

where $x$ is the position along a defect subjected to a force $f = f_S + f_D$. $T$ is a constant related to line tension.

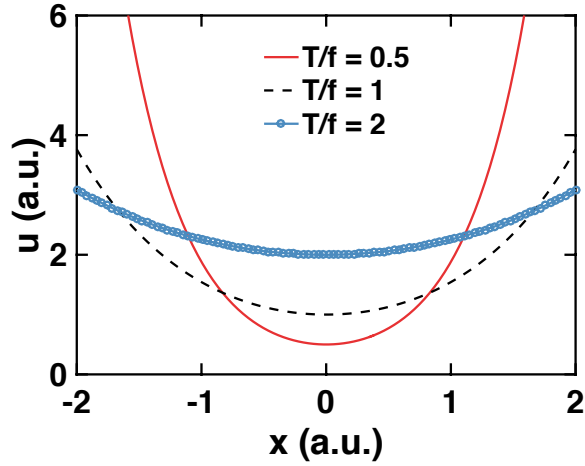

Supplemental Figure S1. Solutions to the catenary cable problem for different values of $T/f$.

For simplicity, we work with the average displacement $u_{avg}$ between two pinning sites spaced $L$ apart.

$$u_{avg} = \frac{1}{L}\int_{-\frac{L}{2}}^{\frac{L}{2}} u(x)\, dx = 2\frac{T^2}{(f_S + f_D)^2 L} \sinh\left(\frac{(f_S + f_D) L}{2T}\right) \tag{30}$$

Using this expression for the average displacement and substituting into Eq. (28), we find

$$\frac{dP}{dt} = C\,\mu \left(\frac{T}{(f_S + f_D)^2} \cosh\left(\frac{(f_S + f_D) L}{2T}\right) - 4\frac{T^2}{(f_S + f_D)^3 L} \sinh\left(\frac{(f_S + f_D) L}{2T}\right)\right)\frac{df_D}{dt} \tag{31}$$

With the approximation $\frac{df_D}{dt} \approx \omega\, f_D$ where $\omega$ is the oscillatory frequency of the experiment, Eq. (31) can be written as

$$\frac{dP}{dt} = C\mu\left(\frac{T}{(f_S+f_D)^2}\cosh\left(\frac{(f_S+f_D)L}{2T}\right) - 4\frac{T^2}{(f_S+f_D)^3 L}\sinh\left(\frac{(f_S+f_D)L}{2T}\right)\right)\omega f_D \quad (32)$$

From this expression, you can imagine doing two experiments: measuring $\frac{dP}{dt}$ as a function of $f_S$ for a fixed $f_D$ and measuring $\frac{dP}{dt}$ as a function of $f_D$ for a fixed $f_S$. Setting $C = \mu = T = L = \omega = 1$, these two experiments would yield polarization responses shown in Figure S2.

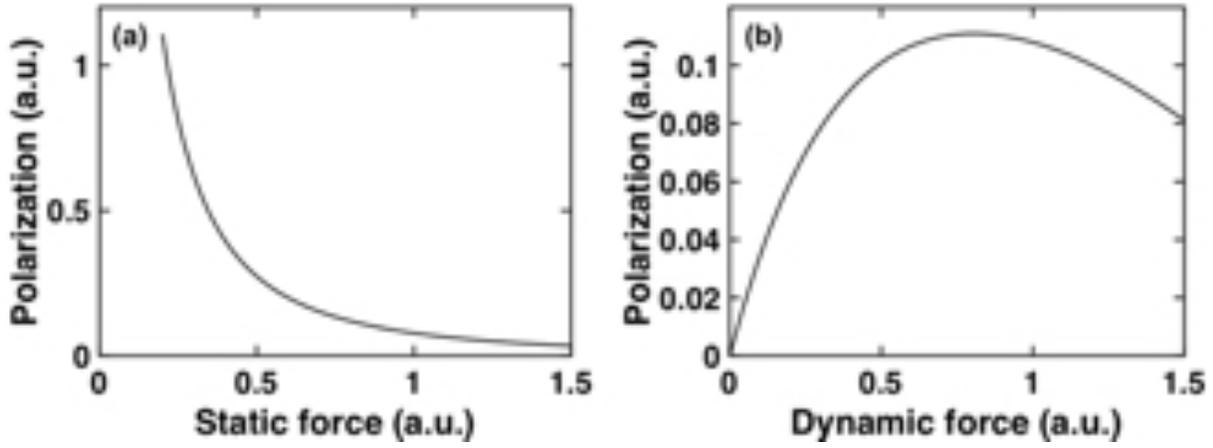

Supplemental Figure S2. Flexoelectric polarization of a pinned defect modeled as a catenary cable as (a) a function of static force for a fixed dynamic force and (b) a function of dynamic force for a fixed static force.

The polarization response shown in Figure S2(a) is directly comparable to the experimental conditions used to acquire the results shown in Figure 3(b) of the main text. Similarly, the polarization response shown in Figure S2(b) is directly comparable to the experimental conditions used to acquire the results shown in Figure 4 in the main text. In both cases, there is good agreement between the catenary cable deformation model for TBs and our experimental results.

In the main text, the fit in Figure 3(b) is of the form

$$f(x) = A\left(\frac{T}{(x+f_D)^2}\cosh\left(\frac{(x+f_D)L}{2T}\right) - 4\frac{T^2}{(x+f_D)^3 L}\sinh\left(\frac{(x+f_D)L}{2T}\right)\right) \quad (33)$$

with fit parameters $A = -1.409$ nC/m$^2$, $T = 1.962$ N m, and $L = 1.589$e-6 m. The experimental value of $f_D = 0.4$ N is used. Our findings suggest that the static force dependency and non-linear flexoelectricity are direct consequences of the non-linear elastic response of pinned TBs.

### S6. Optical characterization

Before and after FxE characterization, TB microstructures were characterized using polarized optical microscopy. Reflection and transmission mode images were acquired on an Olympus PMG 3 and a Nikon SMZ 1500, respectively. A Nomarski prism was used to enhance contrast in the reflection geometry. Because there were no qualitative differences between the transmission and reflection images, only the latter are included in this paper. After FxE characterization, gold electrodes were removed using aqua regia to allow for the imaging of the twin microstructure. In general, imaging before and after *room temperature* FxE characterization on all samples indicated no permanent change in the TB microstructure, suggesting ferroelastic switching cannot explain the room temperature observations. On the contrary, imaging before and after the *temperature-dependent* FxE characterization showed changes to the TB microstructure arising from TB motion and annihilation.

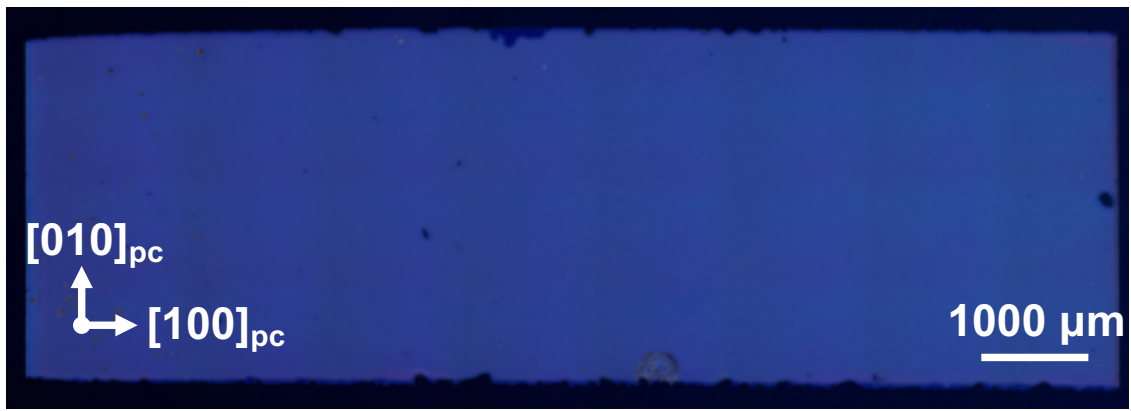

Supplemental Figure S3. Reflection polarized optical microscopy image confirming the lack of twins in the sample referred to as "twin-free" in the manuscript.

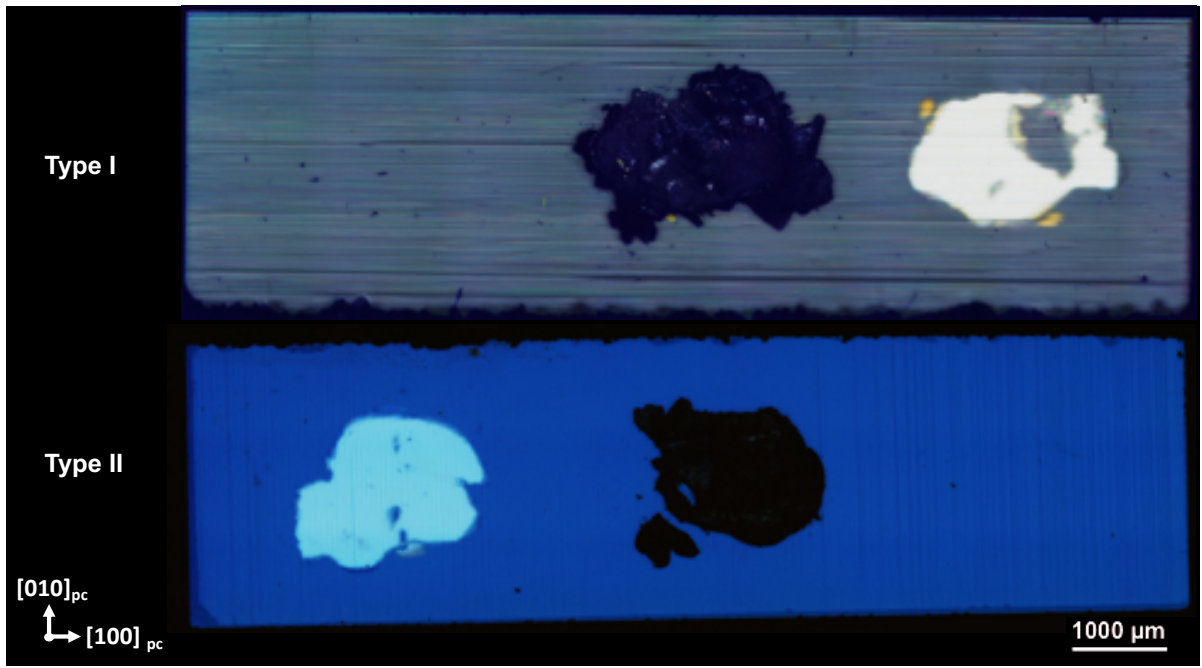

Supplemental Figure S4. Imaging of Type I and II samples after FxE characterization using reflection polarized optical microscopy confirms no permanent changes to the microstructure. Both samples still have uniform lamellar TB microstructures. The splotches correspond to left-over electrode.

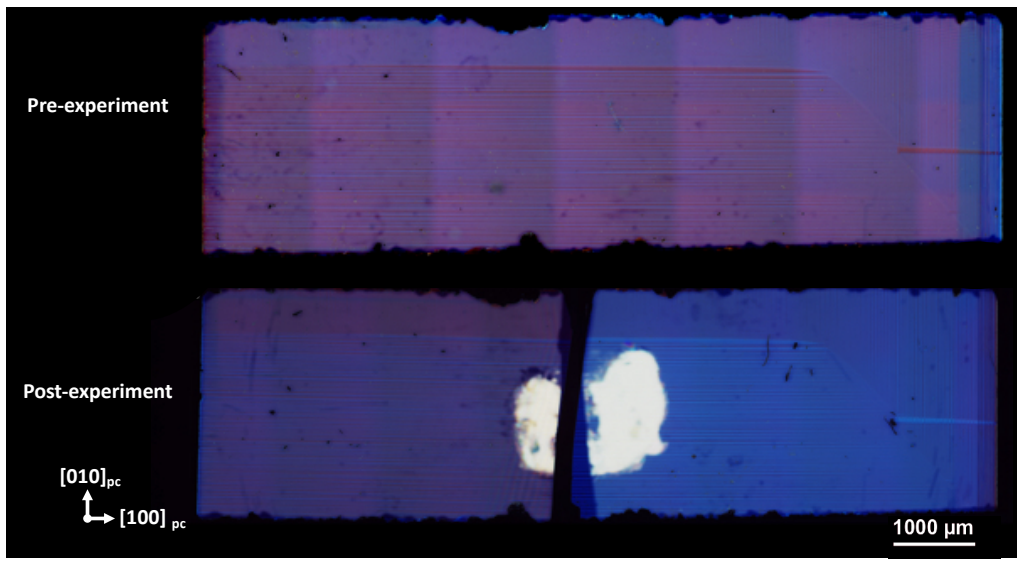

Supplemental Figure S5. Imaging the sample with a mixture of Type I and II TBs before and after FxE characterization using reflection polarized optical microscopy confirms no permanent changes to the microstructure. The vertical lines in the pre-experiment image are imaging artifacts. The central bright region in the post-experiment image is left-over electrode. There is a black portion in the middle of the sample in the post-experiment image because the sample fractured into two pieces after the experiment was performed.

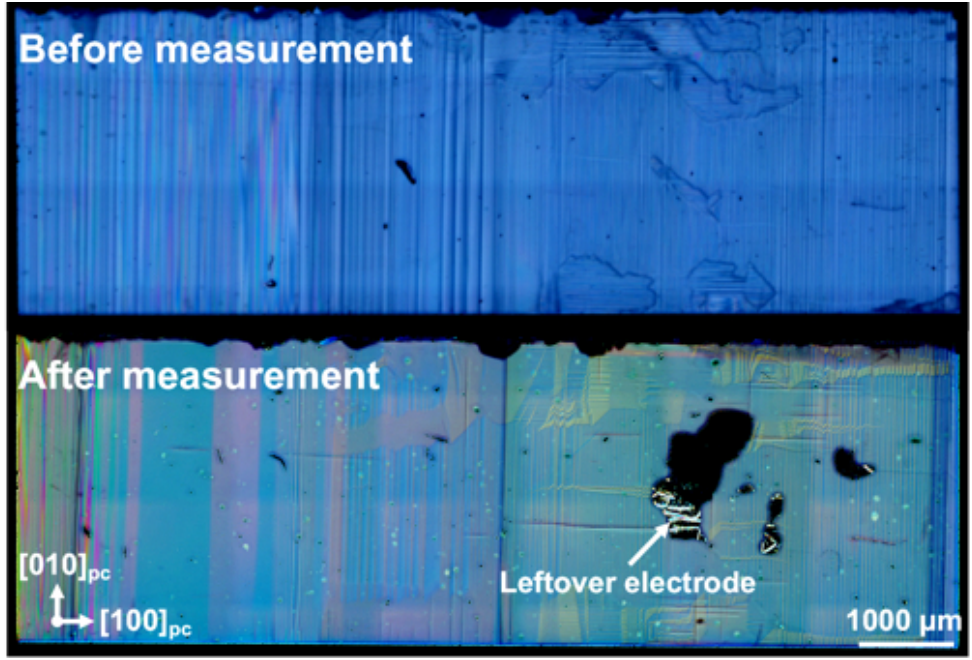

Supplemental Figure S6. Imaging of a sample before and after temperature-dependent FxE characterization using reflection polarized optical microscopy confirms changes to the microstructure.

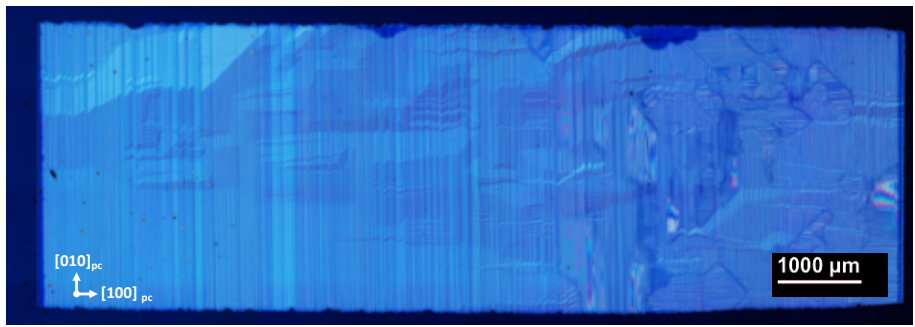

Supplemental Figure S7. Imaging of a sample before the experiment varying the static force. Reflection polarized microscopy confirms the presence of a mixture of TB orientations.

**S7. Additional measurements of anelastic flexoelectricity.**

As described in the manuscript, an anelastic FxE response was observed in samples with TBs. Below we provide additional measurements demonstrating the time-dependent FxE response in the sample with Type I TBs.

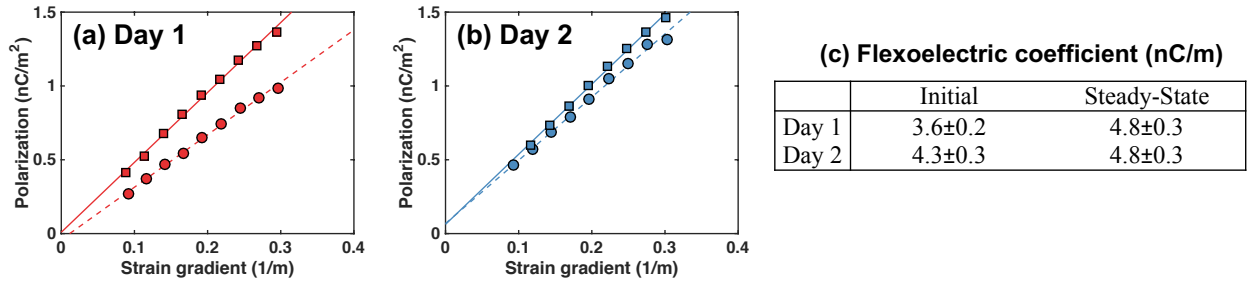

Supplemental Figure S8. Measurements performed on subsequent days on the Type I sample. In both (a) and (b) circles and dashed lines correspond to the initial measurement data and linear fit, respectively. Similarly, squares and solid lines correspond to the steady-state measurement data after strain gradient cycling and linear fit, respectively. The presence of a large anelastic increase in the FxE response on Day 1 is clearly shown in (a). On the second day of measurement, there was a slight recovery of the initial FxE response followed by an anelastic increase to the same steady-state FxE response measured on Day 1 after strain gradient cycling. The values of the slopes are provided in (c). Uncertainties correspond to the 95% confidence interval of the linear fit.

## S8. Sample-to-sample variation and the possibility of piezoelectric contributions

Previous FxE studies have noted sample-to-sample variation in measurements made on nominally the same material (e.g. [1,14]). To avoid similar issues and facilitate quantitative comparison, the Type I (Figure 2, main text), Type II (Figure 2, main text), and mixed Type I+II (Figure 4, main text) samples were cut from the same crystal (they have approximately the same point defect concentration), have the same crystallographic orientation (the bulk FxE response consists of the same combination of tensor components), and were prepared simultaneously (any surface contribution should be similar). All other samples used were cut in the same crystallographic orientation and prepared in as similar conditions as possible.

It is impossible to rule out piezoelectric contributions from TBs and surfaces, however one would anticipate that any piezoelectric polarization should largely cancel because three-point bending subjects samples to equal amounts of tensile and compressive strain.